# Remarkably strong chemisorption of nitric oxide on insulating oxide films promoted by hybrid structure

Zhenjun Song[a], Bin Zhao[a], Hu Xu[b], and Peng Cheng[a]

[a] *Department of Chemistry, Key Laboratory of Advanced Energy Material Chemistry (MOE) and Collaborative Innovation Center of Chemical Science and Engineering (Tianjin), Nankai University, Tianjin 300071, P. R. China.*

[b] *Department of Physics, South University of Science and Technology of China, Shenzhen 518055, P. R. China.*

**ABSTRACT:** The remarkably strong chemical adsorption behaviors of nitric oxide on magnesia (001) film deposited on metal substrate have been investigated by employing periodic density functional calculations with Van der Waals corrections. The molybdenum supported magnesia (001) show significantly enhanced adsorption properties and the nitric oxide is chemisorbed strongly and preferably trapped in flat adsorption configuration on metal supported oxide film, due to the substantially large adsorption energies and transformation barriers. The analysis of Bader charges, projected density of states, differential charge densities, electron localization function, highest occupied orbital and particular orbital with largest Mg-NO-Mg bonding coefficients, are applied to reveal the electronic adsorption properties and characteristics of bonding between nitric oxide and surface as well as the bonding within the hybrid structure. The strong chemical binding of nitric oxide on magnesia deposited on molybdenum slab offers new opportunities for toxic gas detection and treatment. We anticipate that hybrid structure promoted remarkable chemical adsorption of nitric oxide on magnesia in this study will provide versatile strategy for enhancing chemical reactivity and properties of insulating oxide.

## 1. Introduction

Nitric oxide is well known to takes part in various elementary steps which are of special importance and interest for investigating soil chemistry[1, 2], photochemistry[3], catalysis reactions[4, 5], and heterogeneous processes[2]. The discovery of bioregulatory functions of nitric oxide is astounding and revised scientists' understanding of how cells communicate and defend themselves[6]. Nitric oxide can serve as messenger (signaling molecule) in biological processes in immune and nervous systems at low concentration[7-12]. As a by-product of high temperature combustion and exhaust gases generated by motor cars, this colorless gas is one of most common and primary air pollutants. The efficient removal and detection of harmful nitric oxide is still challenging[13] because this small molecule diffuses quickly,[14] and its concentration in biological media changes on time scales of several seconds.[15]

Much attention has been devoted to geometric, catalytic and electronic properties of two-dimensional ionic oxide surfaces such as magnesia (001).[16-24] Magnesia-supported nanoparticles are widely employed as nanocatalyst and the interplay between shape, chemical composition and reactive activity is essential for understanding the structures and dynamics in heterogeneous catalysis processes.[25, 26] The activities and catalytic behaviors of magnesia are frequently dominated and facilitated by the introduction of point defects or unsaturated ions, due to the pronounced charge transfer at low-coordinated O anions acting as good electron source and Mg cations acting as localized hole.[27, 28] Recently, the reverse deposition of thin oxide films on transition metal slabs has emerged as another category of nanocatalyst, which offer new opportunities and interesting routes for heterogeneous catalyst design.[29] For example, for the catalytic adsorption and growth of nanoparticles, Nilius et al investigated the chemical adsorption and self-organization of gold adatoms on FeO films supported by Pt (111) by low temperature scanning tunneling microscopy and spectroscopy.[30] Freund et al summarized the adsorption behaviors toward gold atoms and nanoclusters on thin oxide films grown on metal single crystals.[31] Pacchioni and Freund reviewed the electron transfer at oxide ultrathin films, which is a fundamental process and of significant importance for radical formation in related chemical reactions.[20] Due to the spontaneous formation of superoxide dioxygen anions, greatly enhanced catalytic activity toward oxidation of carbon monoxide was experimentally observed on thin oxide films deposited on metals.[32, 33] Landman et al reported the controlled ethylene hydrogenation on Pt clusters soft-landed on magnesia supported by molybdenum.[34] The dihydrogen adsorption and dissociation on magnesia (001) thin films were researched by Chen et al, and the heterolytic and homolytic splitting of dihydrogen are highly dependent on the choice of the support and the surface morphology (defects and under-coordinated sites).[35, 36] The water molecule dissociates at low activation barrier or even barrierlessly on thin magnesia films.

The adsorption of nitric oxide on magnesia (001) have been studied by employing cluster model density functional theory to evaluate the effect of low coordination on the chemical activity toward NO monomers and dimers.[37] On terrace sites of magnesia (001), nitric oxide interact very weakly with surface, which can be categorized as physisorption. The interaction of nitric oxide with unsaturated cations of magnesia surface results in the formation of chemisorbed species.[38] In this study, the remarkably strong chemical adsorption behaviors of nitric oxide on magnesia (001) film deposited on metal substrate have been investigated by employing periodic density functional calculations with Van der Waals corrections. The strong chemical binding of nitric oxide on magnesia deposited on molybdenum slab offers new opportunities for toxic gas detection and treatment. In addition, we anticipate that hybrid structure promoted remarkable chemical adsorption of nitric oxide on magnesia in this study will provide versatile strategy for enhancing chemical reactivity and properties of insulating oxide.

## 2. Methodologies and models

The interaction of nitric oxide with metal supported magnesia films has been studied by employing periodic density functional theory (DFT) calculations with Van der Waals corrections and spin polarizations. The optB88-vdw[39] corrected Perdew-Burke-Ernzerhof (PBE) functional[40] within generalized gradient approximation (GGA) are used to describe exchange and correlation effects, which includes an accurate description of the uniform electron gas, the dispersion effects, correct behavior under uniform scaling and a smoother potential compared with Perdew-Wang-1991 (PW91) functional.[41] Projector augmented wave (PAW)[42] technique is adopted to describe interactions between core and valence electrons. A kinetic energy cutoff of 500 eV is used to expand the Kohn-Sham orbitals.

To model the hybrid surface, we employed a four layer molybdenum (001) slab consisting of 2 × 2 unit cells; each layer contain 16 molybdenum atoms. The lattice constants of bulk molybdenum are calculated to be 3.152 Å, which agrees very well to experimental value 3.147 Å.[43] The two bottom layers of molybdenum are fixed to mimic bulk properties while the other two metal layers and the magnesia film are fully relaxed until all atomic Hellmann-Feynman forces are less than 0.02 eV Å$^{-1}$. For structure relaxation and energy minimization, the first Brillouin zone is sampled using 2×2×1 and 4×4×1 k-point methes, respectively. The successive slabs are separated by a large vacuum gas with the distance of 17 Å. The transformation reaction profiles and barriers are calculated using the climbing image nudged elastic band (CI-NEB) method implemented in VTST code,[44] which is efficient for finding saddle points and the minimum energy path connecting the given initial and final states. The aforementioned electronic structure calculations are performed using Vienna *Ab Initio* Simulation Package (VASP).[45, 46] The adsorption energy of nitric oxide on bulk magnesia (001) is calculated by formula:

$$E_{ad} = E(NO/MgO(001)) - E(NO) - E(MgO(001)) \qquad (1)$$

The adsorption energies of nitric oxide on metal supported MgO film is obtained as:

$$E_{ad} = E(NO/MgO(001)/Mo(001)) - E(NO) - E(MgO(001)/Mo(001)) \qquad (2)$$

The negative sign of $E_{ad}$ indicates an exothermic adsorption process. After the core charge and valence charge are summed, we use the program developed by Henkelman et al[47-49] to calculate and analyze Bader charge population under very fine fast fourier transform grids. The Visual Molecular Dynamics (VMD) program[50] together with the VESTA program[51] is used to visualize the obtained electronic and geometric structures.

## 3. Results and discussion

### 3.1 Physisorption of nitric oxide on MgO surface

Three typical structures for nitric oxide adsorbing on MgO(001) surface are shown in Figure 1, and the corresponding adsorption energies and geometric parameters are encompassed in Table 1. The other three configurations, with oxygen pointing toward MgO surface or with nitric oxide adsorbing on surface oxygen, show higher relative energies and very larger O(nitric oxide)-$Mg_s$ and N-$O_s$ distances (Figure 2 and Table 2), because of very weak interaction. The adsorption energies for structures in Figure 1 are -312 meV, -273 meV and -267 meV for bridge, 'top of Mg', and oblique configurations respectively, which can be classified as physical adsorption interaction. The adsorption energies determined by orientation and position show very small differences between three typical adsorption configurations. The geometric parameters for the most stable adsorption configuration, second stable adsorption configuration and the third stable adsorption configuration are listed in Table 1. The nitrogen of NO in bridge configuration links with two magnesium, which should account for its slightly higher adsorption energy. The shortest N-Mg bond distances are 2.585 Å, 2.307 Å and 2.414 Å for bridge, 'top of Mg', and oblique configurations, respectively. The N-O distances are only slightly lengthened compared with molecular nitric oxide (1.167 Å). At the adsorption sites, the four Mg1-O and four Mg2-O ionic bonds show largest differences for the bridge configuration. Due to the very weak physisorption strength, the surface alterations are very small with rumpling values 0.047 Å – 0.060 Å. The oblique configuration can be seen as the intermediate isomer between the most stable bridge configuration and the second stable 'on top of Mg' configuration. The minimum energy pathway for nitric oxide diffusing from bridge configuration to 'on top of Mg' configuration is shown in Figure 3. The transformation process show negligible small energy barrier 42 meV, indicating very easy translation of nitric oxide on pristine MgO(001) surface.

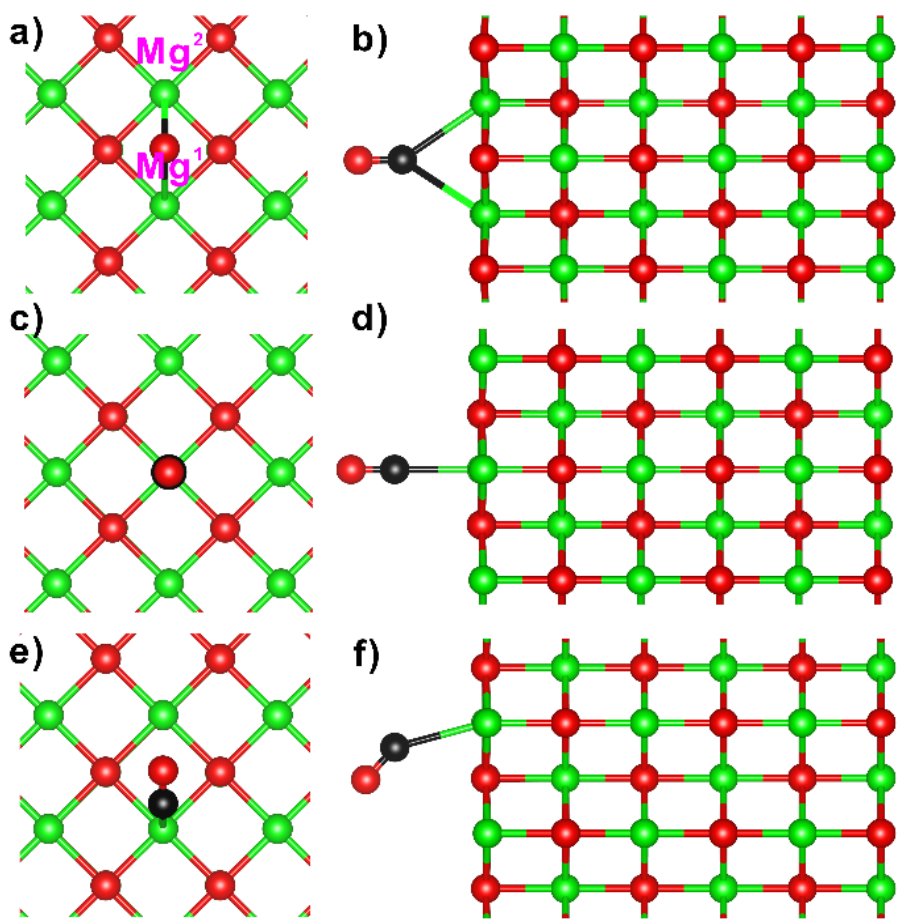

**Figure 1.** The adsorption structures for nitric oxide on MgO(001) surface. (a) (b) The top view and side view of most stable adsorption configuration with nitrogen linked to two surface magnesium atoms (bridge configuration); (c) (d) The top view and side view of the second stable adsorption configuration with nitrogen on top of Mg ('on top of Mg' configuration); (e) (f) The top view and side view of the third stable adsorption configuration with nitrogen adsorbing obliquely on surface magnesium (oblique configuration); The red, green and black balls correspond to oxygen, magnesium and nitrogen atoms.

**Table 1.** The adsorption energies (in eV), geometric parameters (in angstrom) and surface rumpling of the first layer for the most stable bridge adsorption configuration, second stable "top of Mg" adsorption configuration and the third stable oblique adsorption configuration for nitric oxide on pristine MgO (as shown in Figure 1).

|  | Bridge | Top of Mg | Oblique |
|---|---|---|---|
| Adsorption energy | -0.312 | -0.273 | -0.267 |
| N-Mg1 distance | 2.590 | 2.307 | 2.414 |

| | | | | | | |
|---|---|---|---|---|---|---|
| N-Mg2 distance | 2.585 | | —— | | —— | |
| N-O distance | 1.177 | | 1.172 | | 1.174 | |
| Mg1-O distances | 2.145 | 2.145 | 2.129 | 2.129 | 2.136 | 2.136 |
| | 2.118 | 2.118 | 2.129 | 2.129 | 2.125 | 2.125 |
| Mg2-O distances | 2.147 | 2.147 | 2.125 | 2.125 | 2.124 | 2.124 |
| | 2.122 | 2.122 | 2.120 | 2.120 | 2.113 | 2.113 |
| $\Delta z^a$ | 0.053 | | 0.047 | | 0.060 | |

[a] The surface rumpling corresponds to the vertical spacing of the uppermost ion and the most bottom ion, $\Delta z = z_{max}(O/Mg) - z_{min}(O/Mg)$.

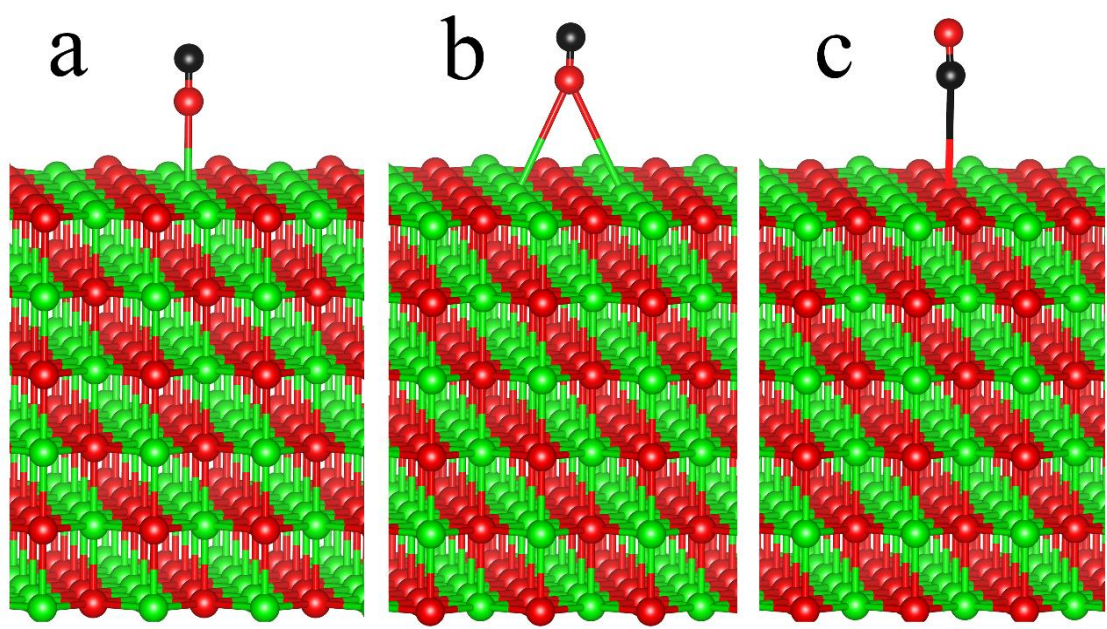

**Figure 2.** The unstable structures with higher adsorption energies for nitric oxide adsorbed on MgO(001) surface. (a) "On top of Mg" configuration, with O pointing toward surface Mg; (b) Bridge configuration, with O pointing toward surface Mg; (c) "On top of O" configuration.

**Table 2.** The adsorption energies (in eV), geometric parameters (in angstrom) for nitric oxide on pristine magnesia with unstable adsorption geometries (a), (b) and (c) (as shown in Figure 2).

| Configuration | (a) | (b) | (c) |
|---|---|---|---|
| Adsorption energy | -0.156 | -0.103 | -0.086 |
| N-O distance | 1.170 | 1.168 | 1.167 |
| O-Mg$_s$ distance | 2.535 | 3.540<br>3.572 | —— |
| N-O$_s$ distance | —— | —— | 3.511 |

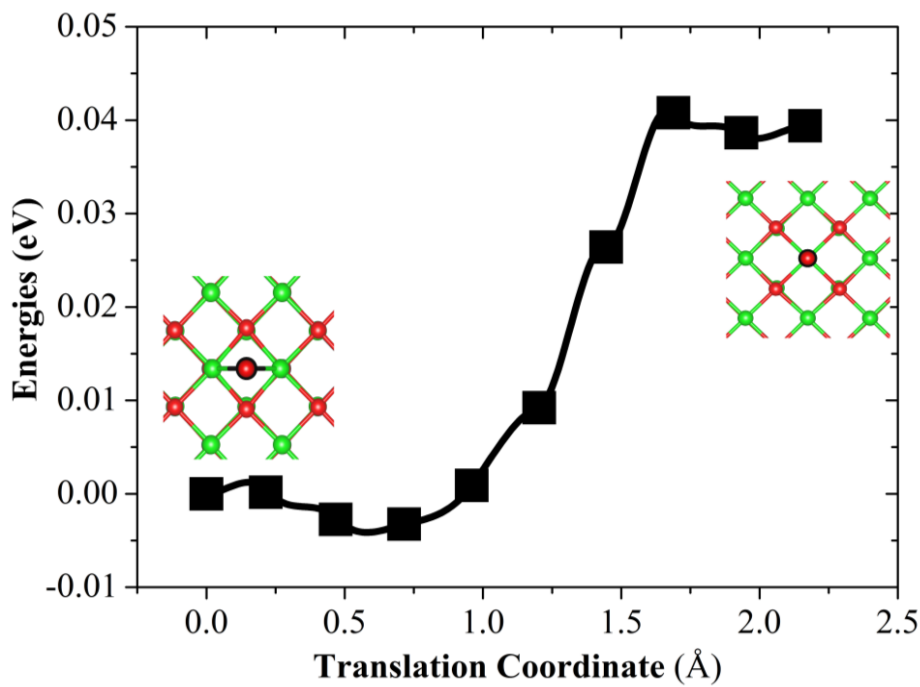

**Figure 3.** Potential energy profile for diffusing reaction from bridge configuration to "on top of Mg" configuration. Geometries of bridge configuration and "on top of Mg" configuration are illustrated. The solid squares are actual images in the nudged elastic band calculations and the curve is minimum energy pathway obtained from force-based interpolation between the images. The red, green and black balls correspond to oxygen, magnesium and nitrogen atoms.

**3.2 Hybrid structure promoted chemical adsorption of nitric oxide**

Both the catalytic activity and stability are of significant importance for catalyst design. Stable catalyst with firm chemical bonds between constituent parts can be prepared at large amount for technological applications. To examine the chemical firmness of the metal supported oxide films, we first calculate the interaction potential between 2 ML MgO film and the molybdenum substrate, by employing climbing-image nudged elastic band method (Figure 4). The combination of Mo substrate and MgO are verified to be smooth process and highly exothermic by -23.785 eV, -20.789 eV and -21.194 eV for 1 – 3 ML MgO respectively, indicating formation of strong chemical interaction between metal substrate and oxide film.

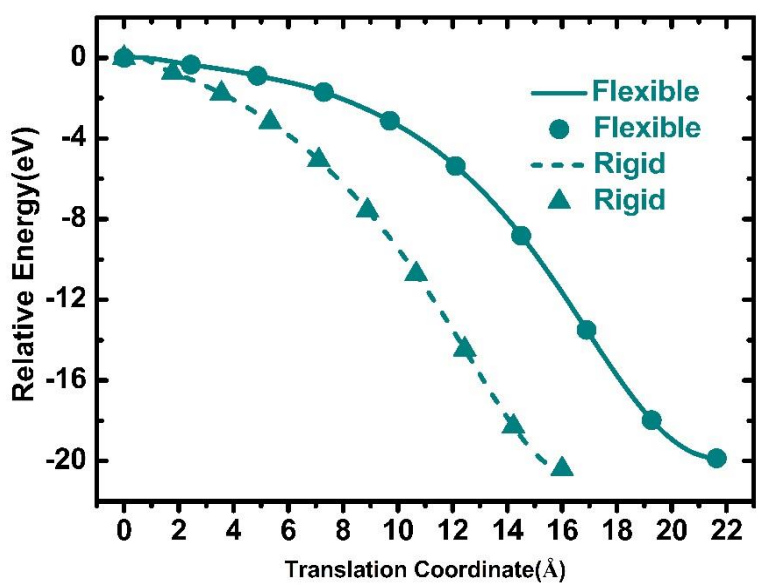

**Figure 4.** Potential energy curve for 2 ML MgO approaching Mo substrate at PBE level. During the flexible translation, the structural geometries of MgO and Mo substrate are optimized. During the rigid translation, the geometries of MgO and Mo substrate are fixed to the optimized MgO/Mo hybrid structure. At the initial translation point, the oxide film are 4.2 Å away from the metal substrate.

The optimized structures for nitric oxide adsorbed on 2 ML MgO(001)/Mo(001) are shown in Figures 5 and 6. The structures in Figure 6 are significantly unstable, with very large adsorbate-Mgs distances and low adsorption energies (Table 5). We mainly focused on the favorable adsorption configurations shown in Figure 5. The structure parameters and adsorption energies for nitric oxide on 1 – 3 ML MgO(001)/Mo(001) are listed in Tables 3 and 4. For all adsorption systems, flat configurations are the most stable adsorption structures with large adsorption energies -1.650 eV, -1.549 eV and -1.303 eV for nitric oxide adsorbed on 1 – 3 ML MgO(001)/Mo(001), respectively. The adsorption energies for "on top of Mg" configuration and bridge configuration are in the ranges -1.069 eV ~ -1.208 eV and -0.892 ~ -1.130 eV, respectively. Generally the adsorption energies are influenced by the thickness of oxide films. The thicker oxide films tend to generate weaker adsorption strength. For each configurations with different film thicknesses, nitric oxide adsorbing on thinnest oxide films lead to the shortest N-Mg1 distances (2.045 Å,

2.077 Å, and 2.106 Å for flat, top and bridge configurations, respectively). Comparing with flat and bridge configurations, the "top of Mg" configurations always possess shortest N-Mg1 bonds for nitric oxide adsorbing on 1 - 3 MgO(001)/Mo(001). In the respect of structural geometry, this bright contrast is due to the nitric oxide in "top of Mg" configuration is linked to only one surface magnesium and the nitric oxide in flat and bridge configurations are attached to two surface magnesium atoms. The flat adsorption configurations form $O_{nitric-oxide}$-Mg2 bonds with distances 2.070 Å, 2.131 Å and 2.157 Å for nitric oxide adsorbing on 1 – 3 ML MgO(001)/Mo(001), repsecitviely, which increase with film thicknesses and show the similar trends as N-Mg1 bonds. After adsorption, the N-O bonds of molecular nitric oxide (1.167 Å) are lengthened to 1.268 Å, 1.266 Å and 1.255 Å for flat adsorption configuration on 1 – 3 ML MgO(001)/Mo(001), respectively.

During the adsorption processes, the surface structures are distorted. The surface rumpling values are much larger than that on bare magnesia (001) surface and decrease with the increasing of film thickness. For example, the surface rumpling are calculated to be 0.302 Å, 0.278 Å and 0.228 Å for nitric oxide adsorption on 1 – 3 ML MgO(001)/Mo(001) with flat configurations, respectively. Apart from the surface rumpling, the surface bonds are partially broken or strengthened, as shown in Table 4. For the flat adsorption configuration, the largest surface Mg-O bond and the shortest surface Mg-O bond are all located in metal supported monolayer oxide, with large difference of Mg1-$O_s$ (Mg2-$O_s$) distances 0.396 Å (0.357 Å). The differences between Mg-O distances are less pronounced for thicker oxide films. For bridge configuration, the surface bond breaking and strengthening are most significant on thinnest oxide film, and the largest bond length difference between longest ionic bond (2.500 Å) and shortest bond (2.210 Å) is 0.290 Å. Because of the weaker interaction between adsorbate and the surface or the high symmetry of the adsorption structure, the surface bonds of 2 - 3 ML MgO(001)/Mo(001) and surface bonds of "top of Mg" configuration on 1 – 3 ML MgO(001)/Mo(001) are much less affected by the adsorption and the surface bond length differences are smaller than 0.1 Å.

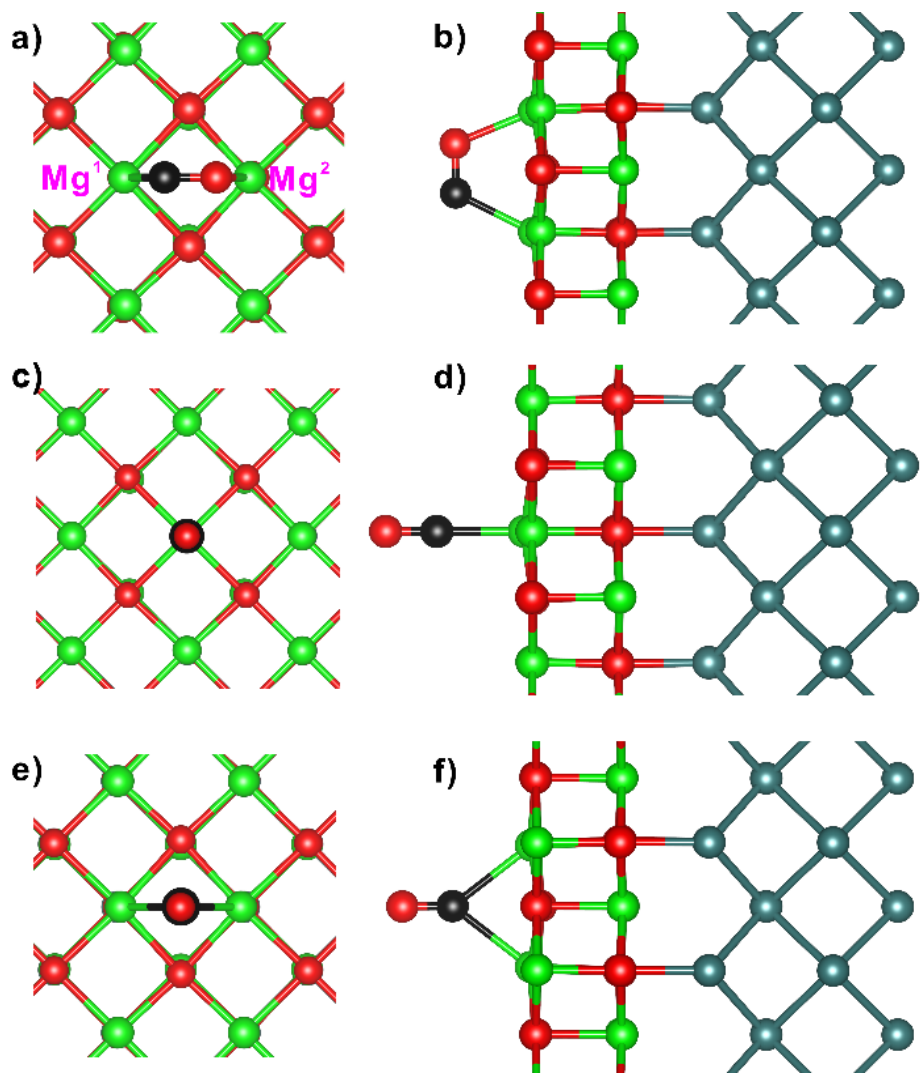

**Figure 5.** The adsorption structures for nitric oxide adsorbed on 2 ML MgO(001)/Mo(001) surfaces. the top view (a) and side view (b) of most stable adsorption configuration with nitrogen and oxygen linked to two surface magnesium atoms (flat configuration); the top view (c) and side view (d) of second stable adsorption configuration with nitrogen adsorbed on one surface magnesium atom (top of Mg configuration); the top view (e) and side view (f) of third stable adsorption configuration with nitrogen linked to two surface magnesium atoms (bridge configuration). The red, green, black and dark cyan balls correspond to oxygen, magnesium, nitrogen and molybdenum atoms.

**Table 3.** The adsorption energies (in eV), geometric parameters (in angstrom) and surface rumpling (in angstrom) of the first layer oxide for the most stable 'flat' adsorption configuration, second stable 'top of Mg' adsorption configuration and the third stable 'bridge' adsorption configuration for nitric oxide on 1 – 3 ML MgO(001)/Mo(001) surfaces.

|  | Flat | | | Top of Mg | | | Bridge | | |
|---|---|---|---|---|---|---|---|---|---|
| Oxide thickness | 1 ML | 2 ML | 3 ML | 1 ML | 2 ML | 3 ML | 1 ML | 2 ML | 3 ML |
| $E_{ad}$ | -1.650 | -1.549 | -1.303 | -1.134 | -1.208 | -1.069 | -1.130 | -1.059 | -0.892 |
| N-Mg1 distance | 2.145 | 2.200 | 2.244 | 2.045 | 2.077 | 2.106 | 2.271 | 2.396 | 2.473 |
| N-Mg2 distance | 2.877 | 2.866 | 2.890 | 3.879 | 3.855 | 3.893 | 2.273 | 2.376 | 2.452 |
| O-Mg2 distance | 2.070 | 2.131 | 2.157 | —— | —— | —— | —— | —— | —— |
| N-O distance | 1.268 | 1.266 | 1.255 | 1.218 | 1.219 | 1.214 | 1.239 | 1.227 | 1.217 |
| $\Delta z^a$ | 0.302 | 0.278 | 0.228 | 0.300 | 0.209 | 0.200 | 0.320 | 0.254 | 0.190 |

$^a$ The surface rumpling corresponds to the vertical spacing of the uppermost ion and the most bottom ion, $\Delta z = z_{max}(O/Mg) - z_{min}(O/Mg)$.

**Table 4.** Surface bond lengths Mg1-$O_s$ and Mg2-$O_s$ at the reaction sites for the most stable 'flat' adsorption configuration, second stable 'top of Mg' adsorption configuration and the third stable 'bridge' adsorption configuration for nitric oxide on 1 – 3 ML MgO(001)/Mo(001) surfaces. All units are set in Å.

| Configuration | Oxide thickness | Mg1-Os distance | | | | Mg2-Os distance | | | |
|---|---|---|---|---|---|---|---|---|---|
| Flat | 1 ML | 2.163 | 2.165 | 2.536 | 2.559 | 2.174 | 2.199 | 2.454 | 2.531 |
|  | 2 ML | 2.268 | 2.270 | 2.341 | 2.343 | 2.279 | 2.282 | 2.325 | 2.328 |
|  | 3 ML | 2.270 | 2.279 | 2.313 | 2.317 | 2.279 | 2.293 | 2.298 | 2.299 |
| Top of Mg | 1 ML | 2.289 | 2.290 | 2.293 | 2.294 | 2.231 | 2.231 | 2.235 | 2.238 |
|  | 2 ML | 2.295 | 2.299 | 2.301 | 2.303 | 2.213 | 2.218 | 2.248 | 2.262 |
|  | 3 ML | 2.287 | 2.288 | 2.289 | 2.296 | 2.217 | 2.221 | 2.247 | 2.254 |
| Bridge | 1 ML | 2.210 | 2.217 | 2.475 | 2.500 | 2.211 | 2.218 | 2.472 | 2.501 |
|  | 2 ML | 2.273 | 2.274 | 2.305 | 2.305 | 2.262 | 2.264 | 2.325 | 2.326 |
|  | 3 ML | 2.252 | 2.253 | 2.295 | 2.315 | 2.257 | 2.272 | 2.288 | 2.294 |

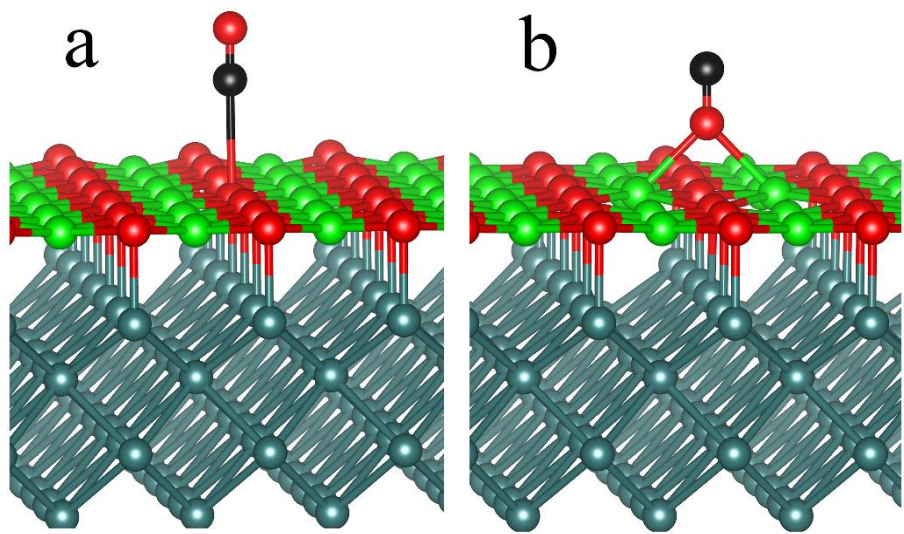

**Figure 6.** The unstable structures with higher adsorption energies for nitric oxide adsorbed on MgO(001) surface. (a) "On top of Mg" configuration, with O pointing toward surface Mg; (b) Bridge configuration, with O pointing toward surface Mg; (c) "On top of O" configuration.

**Table 5.** The adsorption energies (in eV) and geometric parameters (in angstrom) of the first layer oxide for the unstable configurations (a) and (b) (as shown in Figure 6).

| Configuration | (a) | (b) |
| --- | --- | --- |
| Adsorption energy | -0.422 | -0.273 |
| N-O distance | 1.204 | 1.260 |
| O-$Mg_s$ distance | —— | 2.282 / 2.279 |
| N-$O_s$ distance | 2.798 | —— |

In Figure 7, we show the potential energy profile for transformation reaction from flat configuration to "on top of Mg" configuration. The transformation experiences one intermediate state (IM) and two transition states (TS1 and TS2). As listed in Table 4, the energy barrier is 0.692 eV for the transformation from the initial state (flat adsorption configuration) to intermediate state. The nitric oxide at intermediate state further stands up in the subsequent step to form the "on top of Mg" configuration and corresponding activation barrier is 0.515 eV. The barrier heights of transformation processes are much larger than that on bare MgO (001) surface.

During the transformation reaction, the N-O bond lengths fluctuate according to the adsorption energies, as shown in Figure 8. The adsorption energies of nitric oxide at all considered configurations are significantly larger than that on bare MgO (001) surface (as listed in Tables 3). In addition, the adsorption energies of intermediate IM, transition states TS1 and TS2 during configuration transformation are calculated to be -1.158 eV, -0.958 eV, and -0.643 eV, respectively, which are also obviously larger than that on bare oxide surface (as listed in Table 6). Consequently, the toxic gas NO is chemisorbed strongly and preferably trapped in flat adsorption configuration on metal supported oxide film, as a result of the large transformation activation energy and large adsorption energy.

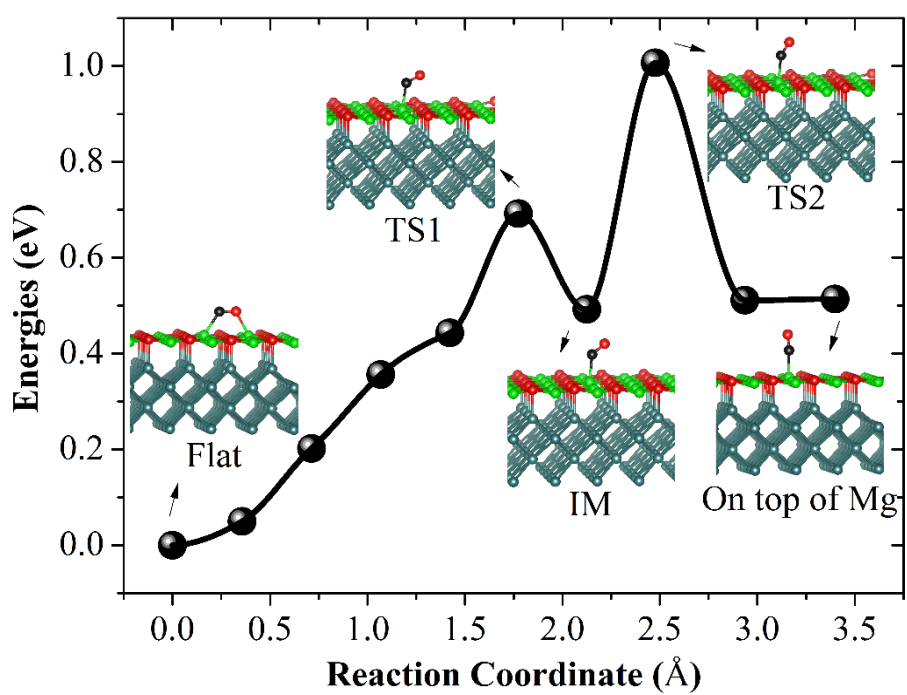

**Figure 7.** Potential energy profile for transformation reaction from flat configuration to "on top of Mg" configuration. Geometries of flat configuration, "on top of Mg" configuration, two transition states (TS1 and TS2), and intermediate state (IM) are illustrated. The solid spheres are actual images in the nudged elastic band calculations and the curve is obtained from force-based interpolation between the images. The red, green, black and dark cyan balls correspond to oxygen, magnesium, nitrogen and molybdenum atoms.

**Table 6.** The N-O distances (in Å), adsorption enerigies ($E_{ad}$, in eV) and activation energies ($E_a$, in eV) of nitric oxide at intermediate state (IM) and transition states (TS1 and TS2).

|  | TS1 | IM | TS2 |
|---|---|---|---|
| N-O | 1.220 | 1.226 | 1.209 |
| $E_{ad}$ | -0.958 | -1.158 | -0.643 |
| $E_a$ | 0.692 | — | 0.515 |

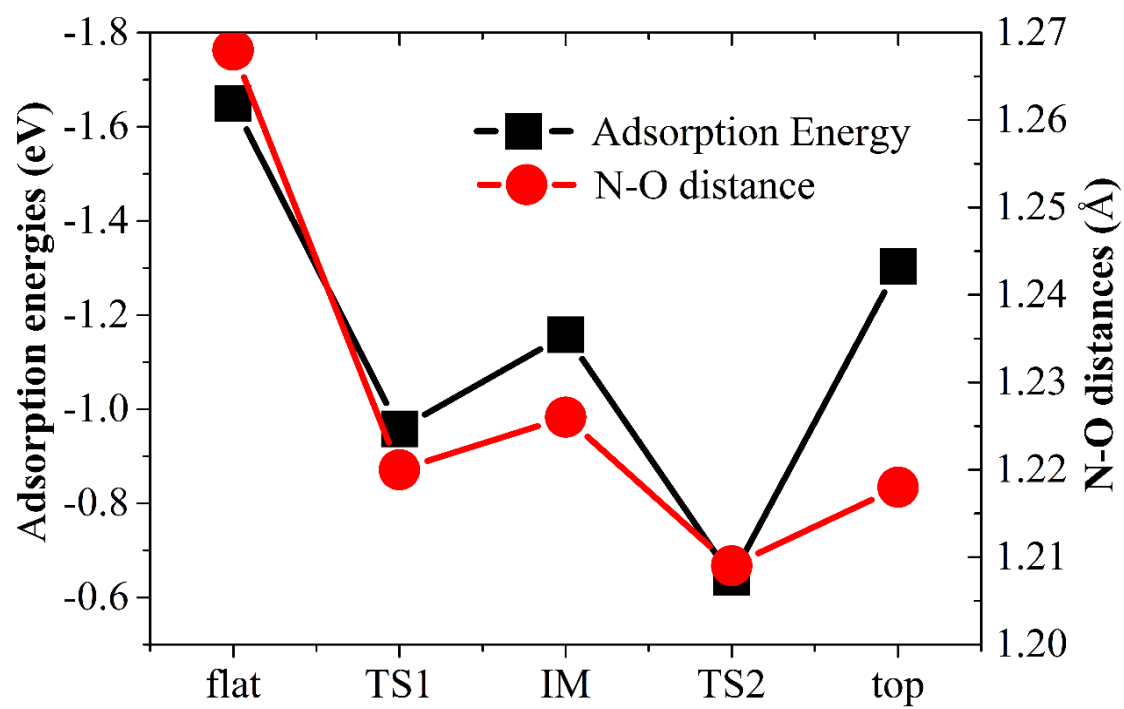

**Figure 8.** Adsorption energies and N-O distances dependent on the configurations during transformation reaction.

### 3.3 Electronic properties

The Bader charge population of NO, Mg1, Mg2, MgO and Mo are calculated, as listed in Table 7, to quantitatively elucidate the electron transfer between adsorbates and hybrid structure. For all the adsorption geometries considered, the adsorbed nitric oxide molecules are negatively charged. The $\pi_{2p}^*$ orbital of negatively charged nitric oxide molecule obtains electrons from the hybrid film, and the bond order of N-O are reduced, which is in accord with the elongated N-O bond distances. The nitric oxide adsorbing on 1 ML MgO(001)/Mo(001) acquire more electrons (-0.865 e, -0.638 e and -0.773 e for flat, top and bridge configurations) than that on thickest 3 ML MgO(001)/Mo(001) (-0.796 e, -0.586 e and -0.609 e for flat, top and bridge configurations), indicating the thicker films are substantially disadvantageous to charge transfer and the electrostatic interaction between nitric oxide and the hybrid films. Nitric oxide in flat configurations gain more electrons than that in bridge and top configuration, which is responsible for the largest adsorption strength in flat configuration. Because of bonding sufficiently with two surface magnesium, the nitric oxide in bridge configurations gain more electrons from the hybrid surface than that in top configurations. Nitrogen atoms with relatively smaller electron affinities acquire less electrons than oxygen atoms of nitric oxide adsorbed on the hybrid structures.

Generally, as the oxide films grow thicker, the Mg1 are more positively charged. Although the coordination environments of Mg1 and Mg2 atoms are significantly dissimilar, the surface magnesium show slightly different positive charges. The magnesia ultrathin films are positively charged with noticeable level (+1.453 e ~ +2.048 e). On the whole, the thinner oxide films with high positive charges are more severely oxidized, leading to the remarkably enhanced activity of monolayer magnesia (001). The molybdenum substrates are negatively charged and serves as the electron reservoir. The molybdenum in NO/1ML MgO(001)/Mo(001) hybrid structures accommodate most electrons, compared with that in NO/2 ML and 3 ML/Mo(001). The molybdenum substrates in flat adsorption configurations, compared with that in top or bridge adsorption configurations, are negatively charged with least amount. These results give strong evidence for the effective charge transfer occurring in flat nitric oxide adsorbed on metal supported ultrathin magnesia, which improve the reactivity of usually inert insulating oxide.

**Table 7. Bader charges (in electron) of NO, Mg1, Mg2, MgO film and Mo substrate.**

| Configuration | Oxide thickness | NO (N, O)[a] | Mg1 | Mg2 | MgO | Mo |

|  | 1 ML | -0.865(-0.140, -0.745) | +1.649 | +1.661 | +2.048 | -1.182 |
| Flat | 2 ML | -0.881(-0.143, -0.738) | +1.674 | +1.684 | +1.649 | -0.767 |
|  | 3 ML | -0.796(-0.086, -0.709) | +1.675 | +1.683 | +1.545 | -0.749 |
|  | 1 ML | -0.638 (-0.138, -0.500) | +1.663 | +1.641 | +1.913 | -1.275 |
| Top of Mg | 2 ML | -0.611(-0.044, -0.567) | +1.680 | +1.663 | +1.529 | -0.918 |
|  | 3 ML | -0.586(-0.039, -0.547) | +1.680 | +1.664 | +1.616 | -1.030 |
|  | 1 ML | -0.773(-0.247, -0.525) | +1.646 | +1.646 | +1.966 | -1.194 |
| Bridge | 2 ML | -0.702(-0.159, -0.543) | +1.673 | +1.674 | +1.570 | -0.868 |
|  | 3 ML | -0.609(-0.125, -0.484) | +1.674 | +1.674 | +1.453 | -0.843 |

[a] The charge values of nitrogen and oxygen of adsorbed nitric oxide are shown in the parentheses.

As shown in Figures 9 - 11, the localized and projected density of states of nitric oxide, surface magnesium, interfacial oxygen and molybdenum are calculated to reveal the electronic properties. Nitric oxide adsorbed with top and bridge configurations occupy states at the Fermi level, while nitric oxide adsorbed on 1 ML MgO(001)/Mo(001)with flat configuration are relatively far away from the Fermi level, indicating the flat configurations is stabilized to lower energy states and should be favored with respect to other adsorption configurations. For the most stable flat adsorption configurations, the relative energy levels of nitric oxide are highly dependent on the film thickness. When the oxide film grow thicker, the states of nitric oxide shift toward higher relative energy levels near Fermi level. Compared with gaseous nitric oxide, the $\pi_{2p}^*$ orbitals of adsorbed nitric oxide with flap and bridge configurations are split due to the presence of coordination environment. However, the $\pi_{2p}^*$ orbitals of adsorbed nitric oxide with top configuration do not branch off near the Fermi level, due to the high coordination symmetry of the nitric oxide. For nitric oxide adsorbing on 1 ML MgO(001)/Mo(001) with flat configuration, the surface magnesium Mg1 and Mg2 occupy states at about -0.6 eV and -1.2 eV below Fermi level, which hybridize well with states peaks of nitric oxide, indicating the effective bonding of Mg1-N and Mg2-O. For "top of Mg" configuration, the electron states of surface magnesium Mg1 and Mg2 locate at rather different energy region, and the DOS curves of Mg2 can be seen as a reference because of no direct bonding interaction with nitric oxide. Mg1 2s states hybridize obviously with nitric oxide at -0.2 eV below Fermi level, due to the strong chemical interaction between Mg1 and nitric oxide. For bridge configuration, the 2s states peaks of surface magnesium hybridize with states of nitric oxide around -1.4 eV and -0.4 eV effectively. At the interface, the $4d_{z^2}$ orbital of interfacial molybdenum and $2p_z$ of interfacial oxygen hybridize broadly, indicating the presence of covalent bonding interaction.

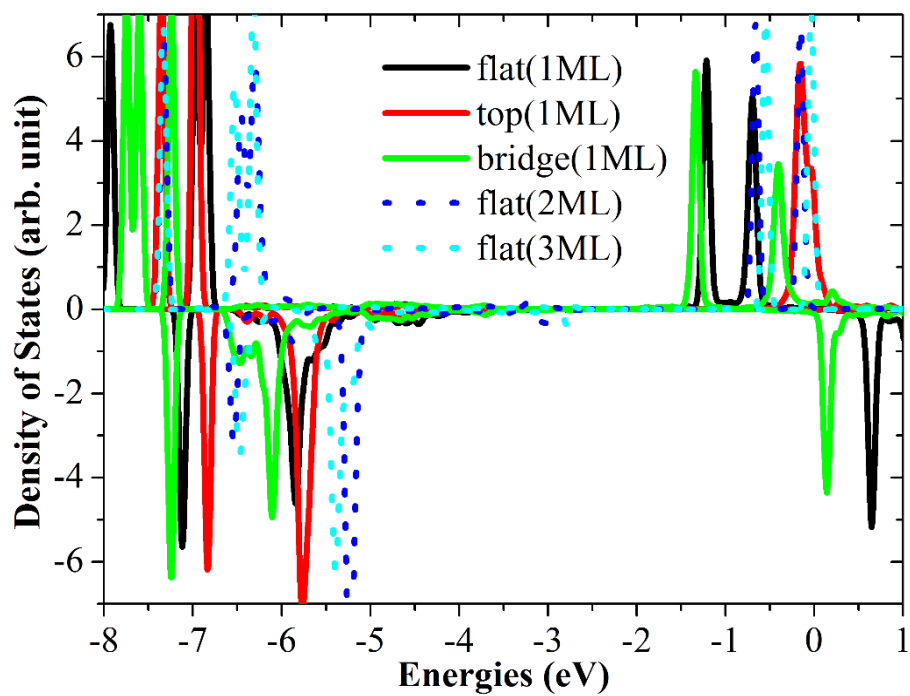

**Figure 9.** Projected density of states of nitric oxide adsorbing on 1 ML MgO(001)/Mo(001) with flat, "on top of Mg" and bridge configurations, respectively; projected density of states of nitric oxide adsorbing on 2 ML and 3 ML MgO(001)/Mo(001) with flat configurations, respectively.

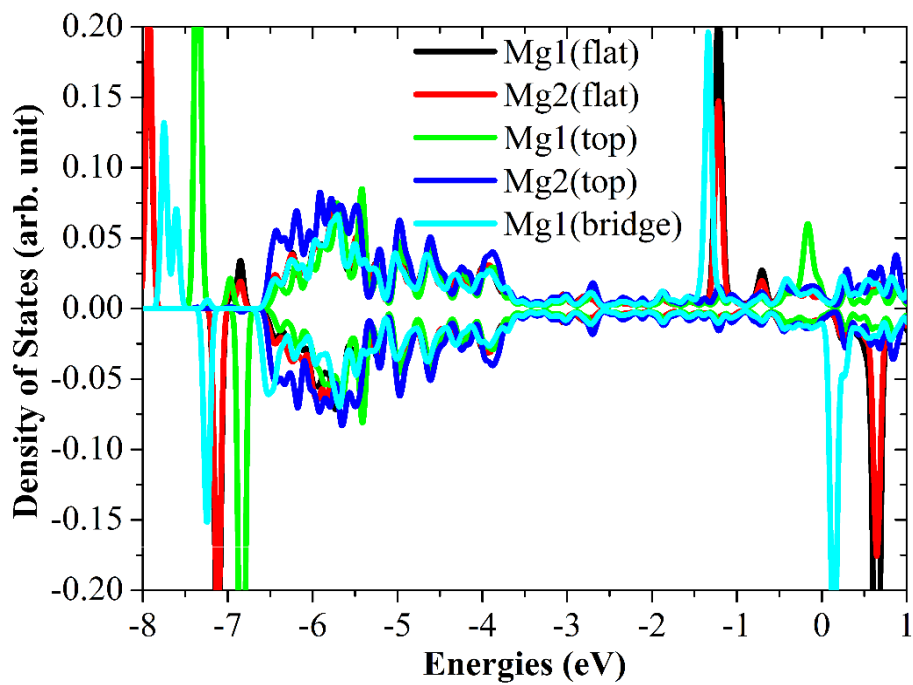

**Figure 10.** Projected density of states of Mg1, Mg2 for nitric oxide adsorbing on 1 ML MgO(001)/Mo(001) with flat, "on top of Mg" and bridge configurations, respectively.

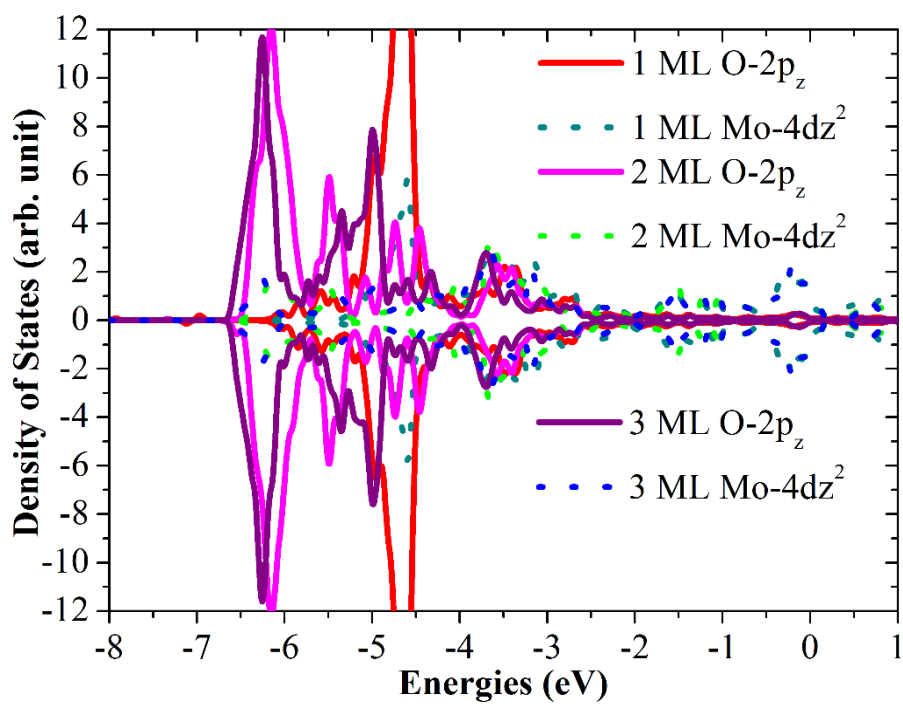

**Figure 11.** Projected density of states of interfacial oxygen and molybdenum for nitric oxide adsorbing on 1 - 3 ML MgO(001)/Mo(001) with flat configuration.

As shown in Figure 12, the differential charge densities contour for nitric oxide adsorbing on 1 ML MgO(001)/Mo(001) with flat configuration display obvious cyan electron depletion densities between N and O of nitric oxide, indicating the N-O bond is substantially weakened, which is in accord with the lengthened N-O bond distance (1.268 Å). The electron depletion phenomenon is also found in the ionic magnesium-oxygen bonds of oxide films, after depositing on molybdenum slab, demonstrating the presence of weakened ionic bonding. More complex charge redistribution occurs at the interfacial area. The interfacial oxygen obtain electrons from interfacial molybdenum. The charge depletion and charge accumulation between interfacial oxygen and molybdenum are right in the z direction, which vividly show the covalent bonding picture and agree well with the significant hybridization effect between Mo-$4dz^2$ and O-$2p_z$ orbitals (Figure 11). Different from the interfacial molybdenum atoms, due to the relative large electron affinity, the molybdenum inside bulk phase generally accumulate electrons, which is responsible for the negative total charge of molybdenum substrate and agrees well with the Bader charge analysis.

The analysis of electron localization function is applied to reveal the nature of bonding between nitric oxide and surface as well as the bonding within the MgO/Mo hybrid structure, as shown in Figure 13. The nitric oxide shows obvious red color, indicating the electron localization effect. The majority of the surface oxygen show great electron localizing ability. In the contrast, the oxygen atoms at the adsorption site show very limited localization effect, probably due to the more significant orbital hybridization of these oxygen atoms with underneath molybdenum (as shown in Figure 14a), which increases the delocalization property of these oxygen atoms.

As displayed in Figure 14, the electronic density of highest occupied molecular orbital[52] mainly spread over molybdenum, which confirms well to the monopolizing large density of states of molybdenum at Fermi level (Figures 9 - 11). From the particular orbital with largest coefficient for Mg-N and Mg-O (N, O from nitric oxide) interaction, we can clearly identify the formation of chemical bonding between nitric oxide and the oxide film (Figure 14b). The oxygen atoms at the adsorption site are different from the rest and show remarkable orbital hybridization with interfacial molybdenum, even at the highest occupied orbital. The spin densities primarily focus on nitric oxide. The oxygen atoms at the adsorption site show tiny spin densities, due to the influence of structure distortion and chemical adsorption.

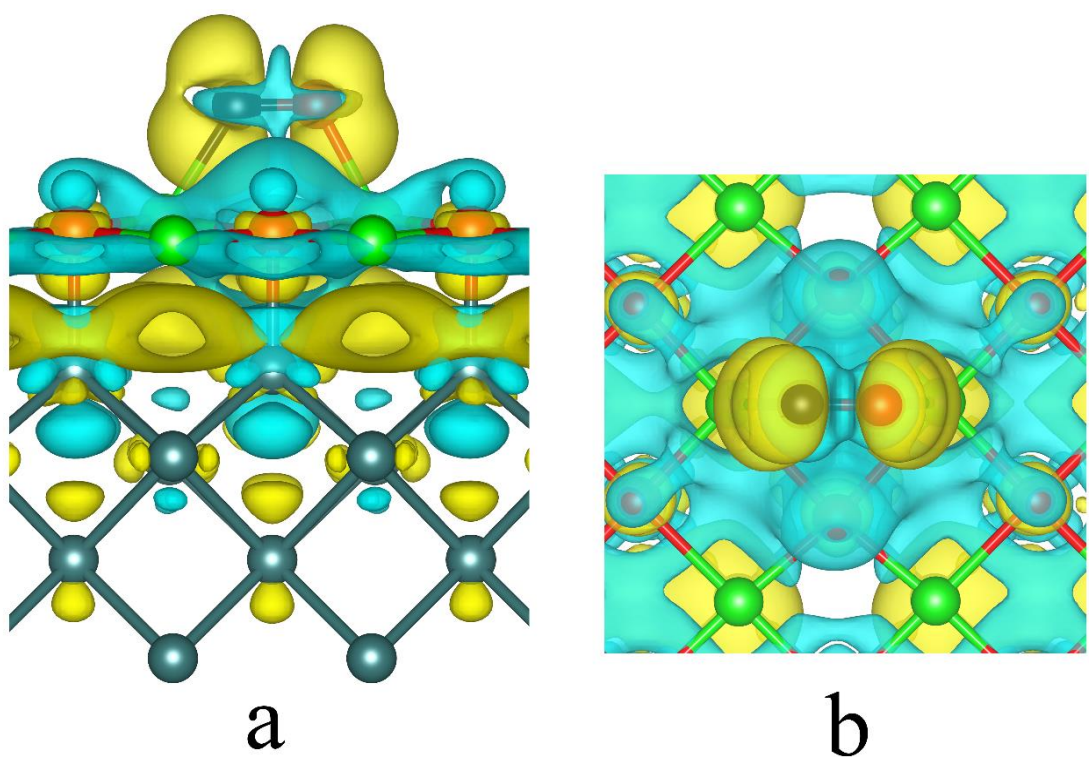

**Figure 12.** Differential charge densities contour for nitric oxide adsorbing on 1 ML MgO(001)/Mo(001) with flat configuration. (a) Side view, (b) Top view. The differential charge densities is defined as $\Delta\rho = \rho(\text{Total}) - \rho(\text{NO}) - \rho(\text{MgO}) - \rho(\text{Mo})$. The isosurface value for the charge densities is set to 0.002 e Bohr$^{-3}$. The yellow and cyan regions correspond to electron accumulation and electron depletion, respectively.

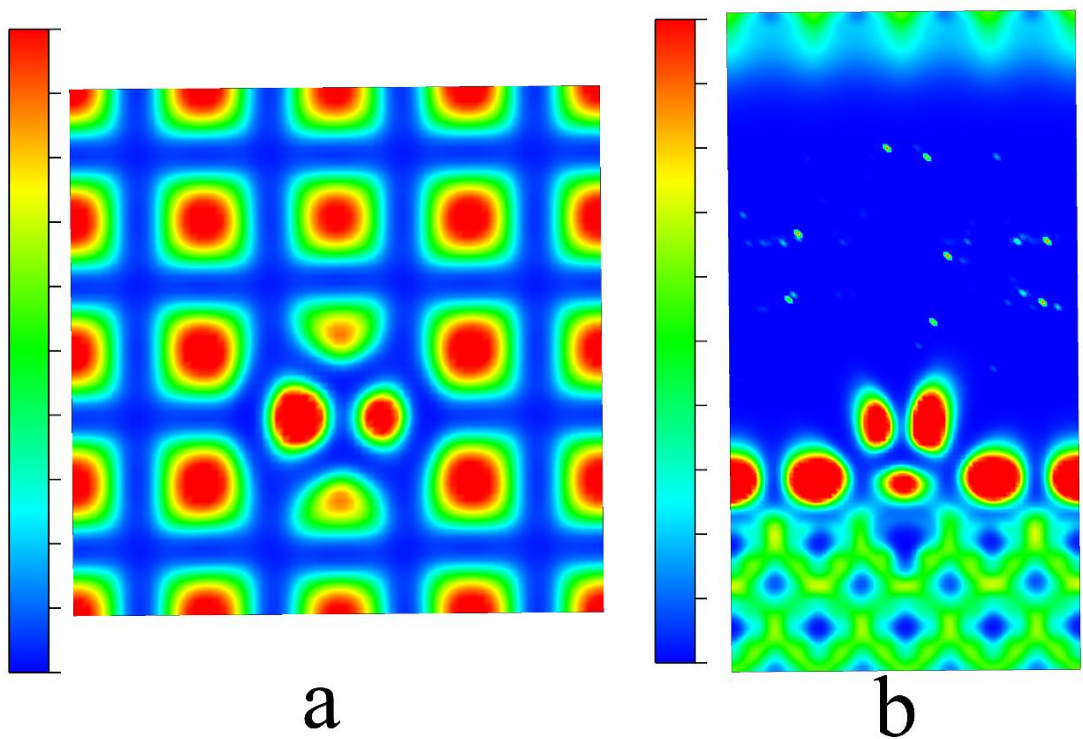

**Figure 13.** Top and side views of electron localization function for nitric oxide adsorbing on 1 ML MgO(001)/Mo(001) with flat configuration. The plots of electron localization function are under the same saturation levels. The blue and red colors correspond to electron delocalization and electron localization, respectively.

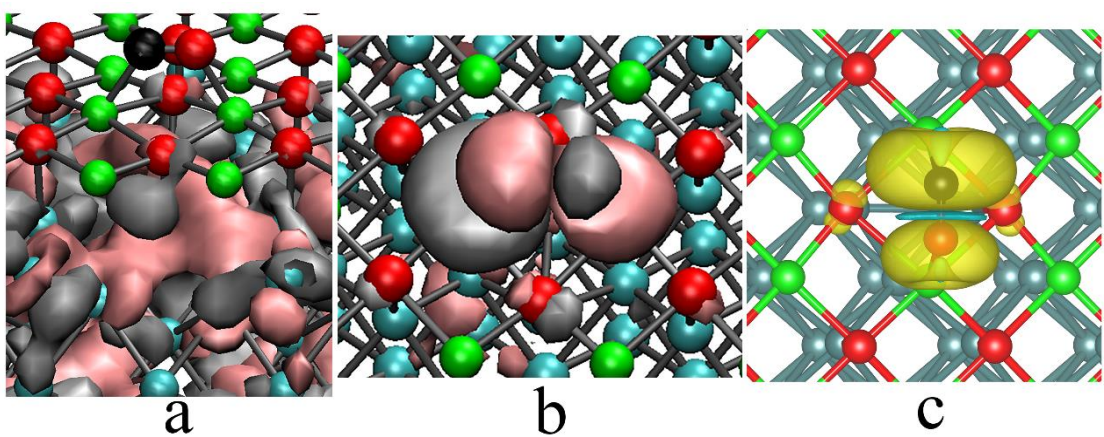

**Figure 14.** Highest occupied molecular orbital (a) and the orbital with largest orbital coefficient for N-Mg and O-Mg bonding interaction (b) and spin density isosurfaces (c) for nitric oxide adsorbing on 1 ML MgO(001)/Mo(001) with flat configuration.

The isosurface values for orbital and spin density are 0.012 e Å$^{-3}$ and 0.001 e Bohr$^{-3}$, respectively.

## 4. Conclusions

In summary, the remarkably strong chemical adsorption behaviors of nitric oxide on magnesia (001) film deposited on metal substrate have been investigated by employing periodic density functional calculations with Van der Waals corrections. Nitric oxide adsorption on bulk magnesia (001) shows very weak physisorption strength. The transformation and desorption processes should be very easy and show negligible energy barrier. The molybdenum supported magnesia (001) show significantly enhanced adsorption properties toward nitric oxide. The flat configuration is most favorable, showing adsorption energies as large as -1.65 eV.

The transformation reaction from most stable flat configuration to second stable "on top of Mg" configuration experiences intermediate state and two transition states. The energy barrier is 0.692 eV for the transformation from the initial state (flat adsorption configuration) to intermediate state. The nitric oxide at intermediate state further stands up in the subsequent step to form the "on top of Mg" configuration and corresponding activation barrier is 0.515 eV. The barrier heights of transformation processes are much larger than that on bare magnesia (001) surface. Consequently, the toxic gas NO is chemisorbed strongly and preferably trapped in flat adsorption configuration on metal supported oxide film, as a result of the large transformation activation energy and large adsorption energy.

Nitric oxide in flat configurations gain more electrons than that in bridge and top configurations, which is responsible for the largest adsorption strength in flat configuration. On the whole, the thinner oxide films with high positive charges are more severely oxidized, leading to the remarkably enhanced activity of monolayer magnesia (001). The molybdenum substrates are negatively charged and serves as the electron reservoir. The molybdenum substrates in flat adsorption configurations, compared with that in top or bridge adsorption configurations, are negatively charged with least amount, giving strong evidence for the effective charge transfer occurring in flat nitric oxide adsorbed on metal supported ultrathin magnesia. Besides Bader charge analysis, the analysis of projected density of states, differential charge densities, electron localization function, highest occupied orbital and particular orbital with largest Mg-NO-Mg bonding coefficients, are applied to reveal the electronic adsorption properties and characteristics of bonding between nitric oxide and surface as well as the bonding within the hybrid structure. The strong chemical binding of nitric oxide on magnesia deposited on molybdenum slab offers new opportunities for toxic gas treatment. We anticipate that hybrid structure promoted remarkable chemical adsorption of nitric oxide on magnesia in this study will provide versatile strategy for enhancing chemical reactivity and properties of insulating oxide.

## Acknowledgements:

This work was supported by the NSFC (Grants 21625103, 21571107, and 21473121), Project 111 (Grant B12015), and the SFC of Tianjin (Grant 15JCZDJC37700). The density functional calculations in this research were performed on TianHe-1(A) at National Supercomputer Center in Tianjin.

## References:

1. R. Oswald, T. Behrendt, M. Ermel, D. Wu, H. Su, Y. Cheng, C. Breuninger, A. Moravek, E. Mougin, C. Delon, B. Loubet, A. Pommerening-Roser, M. Sorgel, U. Poschl, T. Hoffmann, M. O. Andreae, F. X. Meixner and I. Trebs, *Science*, 2013, 341, 1233-1235.
2. K. N. Crabtree, M. R. Talipov, O. Martinez, G. D. O'Connor, S. L. Khursan and M. C. McCarthy, *Science*, 2013, 342, 1354-1357.
3. A. A. Eroy-Reveles, Y. Leung, C. M. Beavers, M. M. Olmstead and P. K. Mascharak, *J Am Chem Soc*, 2008, 130, 4447-4458.
4. S. R. Zhang, L. Nguyen, J. X. Liang, J. J. Shan, J. Y. Liu, A. I. Frenkel, A. Patlolla, W. X. Huang, J. Li and F. Tao, *Nat Commun*, 2015, 6.
5. L. Nguyen, S. R. Zhang, L. Wang, Y. Y. Li, H. Yoshida, A. Patlolla, S. Takeda, A. I. Frenkel and F. Tao, *Acs Catal*, 2016, 6, 840-850.
6. E. Culotta and D. Koshland, *Science*, 1992, 258, 1862-1865.
7. D. Suhag, A. K. Sharma, P. Patni, S. K. Garg, S. K. Rajput, S. Chakrabarti and M. Mukherjee, *Journal of Materials Chemistry B*, 2016.
8. C. Bogdan, *Nat Immunol*, 2001, 2, 907-916.
9. J. MacMicking, Q.-w. Xie and C. Nathan, *Annual review of immunology*, 1997, 15, 323-350.
10. V. Calabrese, C. Mancuso, M. Calvani, E. Rizzarelli, D. A. Butterfield and A. M. G. Stella, *Nature Reviews Neuroscience*, 2007, 8, 766-775.
11. J. E. Yuste, E. Tarragon, C. M. Campuzano and F. Ros-Bernal, *Frontiers in cellular neuroscience*, 2015, 9.
12. C. Vila-Verde, A. Marinho, S. Lisboa and F. Guimarães, *Neuroscience*, 2016, 320, 30-42.
13. M. Strianese and C. Pellecchia, *Coordination Chemistry Reviews*, 2016, 318, 16-28.
14. A. Denicola, J. M. Souza, R. Radi and E. Lissi, *Archives of Biochemistry and Biophysics*, 1996, 328, 208-212.

15. F. Bedioui, D. Quinton, S. Griveau and T. Nyokong, *Phys Chem Chem Phys*, 2010, 12, 9976-9988.
16. L. Giordano, C. Di Valentin, J. Goniakowski and G. Pacchioni, *Phys Rev Lett*, 2004, 92.
17. M. Sterrer, M. Yulikov, E. Fischbach, M. Heyde, H. P. Rust, G. Pacchioni, T. Risse and H. J. Freund, *Angew Chem Int Edit*, 2006, 45, 2630-2632.
18. E. Finazzi, C. Di Valentin, G. Pacchioni, M. Chiesa, E. Giamello, H. J. Gao, J. C. Lian, T. Risse and H. J. Freund, *Chem-Eur J*, 2008, 14, 4404-4414.
19. T. Konig, G. H. Simon, U. Martinez, L. Giordano, G. Pacchioni, M. Heyde and H. J. Freund, *Acs Nano*, 2010, 4, 2510-2514.
20. G. Pacchioni and H. Freund, *Chem Rev*, 2013, 113, 4035-4072.
21. F. Stavale, X. Shao, N. Nilius, H. J. Freund, S. Prada, L. Giordano and G. Pacchioni, *J Am Chem Soc*, 2012, 134, 11380-11383.
22. M. Sterrer, T. Risse, L. Giordano, M. Heyde, N. Nilius, H. P. Rust, G. Pacchioni and H. J. Freund, *Angew Chem Int Edit*, 2007, 46, 8703-8706.
23. F. Donati, S. Rusponi, S. Stepanow, C. Wackerlin, A. Singha, L. Persichetti, R. Baltic, K. Diller, F. Patthey, E. Fernandes, J. Dreiser, Z. Sljivancanin, K. Kummer, C. Nistor, P. Gambardella and H. Brune, *Science*, 2016, 352, 318-321.
24. I. G. Rau, S. Baumann, S. Rusponi, F. Donati, S. Stepanow, L. Gragnaniello, J. Dreiser, C. Piamonteze, F. Nolting, S. Gangopadhyay, O. R. Albertini, R. M. Macfarlane, C. P. Lutz, B. A. Jones, P. Gambardella, A. J. Heinrich and H. Brune, *Science*, 2014, 344, 988-992.
25. K. Rossi, T. Ellaby, L. O. Paz-Borbon, I. Atanasov, L. Pavan and F. Baletto, *J Phys-Condens Mat*, 2017, 29.
26. G. G. Asara, L. O. Paz-Borbon and F. Baletto, *Acs Catal*, 2016, 6, 4388-4393.
27. G. Haas, A. Menck, H. Brune, J. V. Barth, J. A. Venables and K. Kern, *Phys Rev B*, 2000, 61, 11105-11108.
28. S. Abbet, E. Riedo, H. Brune, U. Heiz, A. M. Ferrari, L. Giordano and G. Pacchioni, *J Am Chem Soc*, 2001, 123, 6172-6178.
29. H. J. Freund and G. Pacchioni, *Chem Soc Rev*, 2008, 37, 2224-2242.
30. N. Nilius, E. D. L. Rienks, H. P. Rust and H. J. Freund, *Phys Rev Lett*, 2005, 95.
31. T. Risse, S. Shaikhutdinov, N. Nilius, M. Sterrer and H. J. Freund, *Accounts Chem Res*, 2008, 41, 949-956.
32. Y. N. Sun, L. Giordano, J. Goniakowski, M. Lewandowski, Z. H. Qin, C. Noguera, S. Shaikhutdinov, G. Pacchioni and H. J. Freund, *Angew Chem Int Edit*, 2010, 49, 4418-4421.
33. A. Gonchar, T. Risse, H. J. Freund, L. Giordano, C. Di Valentin and G. Pacchioni, *Angew Chem Int Edit*, 2011, 50, 2635-2638.
34. A. S. Crampton, M. D. Rotzer, C. J. Ridge, F. F. Schweinberger, U. Heiz, B. Yoon and U. Landman, *Nat Commun*, 2016, 7.
35. H. Y. T. Chen, L. Giordano and G. Pacchioni, *J Phys Chem C*, 2013, 117, 10623-10629.
36. Z. J. Song and H. Xu, *Appl Surf Sci*, 2016, 366, 166-172.
37. C. Di Valentin, G. Pacchioni, S. Abbet and U. Heiz, *J Phys Chem B*, 2002, 106, 7666-7673.
38. C. Di Valentin, G. Pacchioni, M. Chiesa, E. Giamello, S. Abbet and U. Heiz, *J Phys Chem B*, 2002, 106, 1637-1645.
39. M. Dion, H. Rydberg, E. Schroder, D. C. Langreth and B. I. Lundqvist, *Phys Rev Lett*, 2004, 92.
40. J. P. Perdew, K. Burke and M. Ernzerhof, *Phys. Rev. Lett*, 1996, 77, 3865-3868.
41. J. P. Perdew, J. A. Chevary, S. H. V. K. A. Jackson, M. R. Pederson, R. Mark, D. J. Singh and C. Fiolhais, *Phys. Rev. B*, 1992, 46, 6671-6687.
42. G. Kresse and D. Joubert, *Phys Rev B*, 1999, 59, 1758-1775.
43. G. R. Stewart and A. L. Giorgi, *Phys. Rev. B*, 1979, 19, 5704-5710.
44. G. Henkelman, B. P. Uberuaga and H. Jonsson, *J Chem Phys*, 2000, 113, 9901-9904.

45. G. Kresse and J. Hafner, *Phys. Rev. B*, 1993, 47, 558-561.
46. G. Kresse and J. Furthmuller, *Phys. Rev. B*, 1996, 54, 11169-11186.
47. W. Tang, E. Sanville and G. Henkelman, *J Phys-Condens Mat*, 2009, 21.
48. E. Sanville, S. D. Kenny, R. Smith and G. Henkelman, *J Comput Chem*, 2007, 28, 899-908.
49. G. Henkelman, A. Arnaldsson and H. Jonsson, *Comp Mater Sci*, 2006, 36, 354-360.
50. W. Humphrey, A. Dalke and K. Schulten, *J. Mol. Graphics*, 1996, 14, 33-38.
51. K. Momma and F. Izumi, *J Appl Crystallogr*, 2008, 41, 653-658.
52. The VASPMO program developed by Yang Wang is used to Visualize wavefunctions.